\documentclass[10pt,journal]{IEEEtran}

\usepackage{ifpdf}
\usepackage{caption}
\usepackage{amssymb}%
\usepackage{pifont}%
  \usepackage{cite}
\usepackage{svg}
\usepackage{hyperref}
\usepackage{algorithm}
\usepackage{algorithmic}

\usepackage{amsmath}
\usepackage{acronym}
\usepackage{array}
\usepackage{mdwmath}
\usepackage{mdwtab}

\usepackage{xcolor}
\usepackage{multicol}
\usepackage{tabularx}
\usepackage{longtable}
\usepackage{multirow}
\usepackage{subfig}
\usepackage{algorithm}
\usepackage{url}
\usepackage{etoolbox}
\usepackage{stackengine}
\usepackage{graphicx}
\usepackage{subcaption}
\usepackage{hhline}
\usepackage{booktabs}
\usepackage{todonotes}

\usepackage{stackengine}
\usepackage{amssymb}

\usepackage{url}

\begin{document}
\title{A Framework to Prevent Biometric Data Leakage in the Immersive Technologies Domain
}
\author{Keshav Sood, Iynkaran Natgunanathan, Uthayasanker Thayasivam, Vithurabiman Senthuran, Xiaoning Zhang, and Shui Yu
\IEEEcompsocitemizethanks{\IEEEcompsocthanksitem 
\IEEEcompsocthanksitem  Keshav Sood,~and Iynkaran Natgunanathan are with the School of Information Technology, Deakin University, Geelong, Australia,  \\ E-mail: keshav.sood@deakin.edu.au, iynkaran.natgunanathan@deakin.edu.au;
\\Uthayasanker Thayasivam and Vithurabiman Senthuran is with the Department of Computer Science and Engineering, University of Moratuwa, Sri Lanka. \\E-mail:rtuthaya@cse.mrt.ac.lk, vithurabimans@cse.mrt.ac.lk;\\
Xiaoning Zhang is with the School of Electronic Science and Engineering. University of Electronic Science and Technology of China, Chengdu, China. E-mail:  xnzhang@uestc.edu.cn;
\\
Shui Yu and is with School of Computer Science, University of Technology Sydney, Australia, E-mail: shui.yu@uts.edu.au;
\\}}

\markboth{IEEE}%
{Shell \MakeLowercase{\textit{et al.}}: Bare Advanced Demo of IEEEtran.cls for IEEE Computer Society Journals}
\IEEEtitleabstractindextext{%
\begin{abstract}
Doubtlessly, the immersive technologies have potential to ease people’s life and uplift economy, however the obvious data privacy risks cannot be ignored. For example, a participant wears a 3D headset device which detects participant’s head motion to track the pose of participant’s head to match the orientation of camera with participant’s eyes positions in the real-world. In a preliminary study, researchers have proved that the voice command features on such headsets could lead to major privacy leakages. By analyzing the facial dynamics captured with the motion sensors, the headsets suffer security vulnerabilities revealing a user’s sensitive speech without user’s consent. The psychography data (such as voice command features, facial dynamics, etc.) is sensitive data and it should not be leaked out of the device without users consent else it is a privacy breach. To the best of our literature review, the work done in this particular research problem is very limited. Motivated from this, we develop a simple technical framework to mitigate sensitive data (or biometric data) privacy leaks in immersive technology domain. The performance evaluation is conducted in a robust way using six data sets, to show that the proposed solution is effective and feasible to prevent this issue.     

\end{abstract}

\begin{IEEEkeywords}
Immersive Technologies, VR/XR technologies, cognitive ability, human rights, biometric data privacy
\end{IEEEkeywords}}

\maketitle

\IEEEdisplaynontitleabstractindextext

\IEEEpeerreviewmaketitle

\ifCLASSOPTIONcompsoc
\IEEEraisesectionheading{\section{Introduction and Background}\label{sec:introduction}}
\else
\section{Introduction}
\label{sec:introduction}
\fi

\IEEEPARstart{T}{he} use of immersive technologies (Augmented Reality (AR), Virtual Reality (VR) or  Metaverse) has been in existence since 1990’s but the use was limited to defence sector~\cite{cipresso2018past}. Recently, the technology has shown its potential to revolutionize many sectors. For example, in the education sector, for example digital humanities, scholars use diverse digital tools ranges from a small mobile phone to as large as a virtual reality lab. In medical sciences, students interact with realistic representations of anatomy without using real patients~\cite{dhar2021augmented}. According to Statista the Australian revenue in the immersive technology market is projected to reach out US\$1,053 million in 2023\footnote{https://www.statista.com/outlook/amo/ar-vr/australia-oceania, assessed on 02/05/2023}. Further, it has a potential to strengthen global economy by US\$1.5 trillion by 2030. Doubtlessly, the technology has potential to ease people’s life and uplift economy, however the obvious data privacy risks cannot be ignored. The immersive technologies are currently poorly developed (in privacy context) and there is a significant gap in the technology which hinders its use in secure ways~\cite{10271790}.   \par 
The immersiveness rely on participants data, but the data harvesting technologies often violates the user privacy rights. 
Particularly, in immersive technologies, a participant wears a 3D headset device which detects participant’s head motion to track the pose of participant’s head to match the orientation of camera with participant’s eyes positions in the real-world~\cite{shi2021face},~\cite{8919581}. This device also has an eye tracking sensor which detects both pupil dilation in response to visual stimuli and cognitive responses leading to increased levels of emotional activation– revealing behavioural sensitive information that a user never intended to reveal~\cite{john2020security}. By analysing the facial dynamics captured with the motion sensors, the headsets suffer security vulnerabilities revealing a user’s sensitive speech without user’s consent. This leaked behavioural (psychography) data can feed into misinformation drivers including social, political, and electoral interference campaigns 
poses threats to national security which costs organisations millions of dollars.
\par 
Let alone the privacy, the collection of biometric data enables profiling of participant(s) without their consent and consequently could adversely impact the psychography of participant, eventually violates human rights. This is the disregard for humanity by such technologies. Researchers emphasized that~\textit{``the immersive nature of AR/VR makes it difficult to mitigate risks by applying existing privacy policies and practices from other digital media. It requires innovative new approaches to transparency, choice, and security."}~\cite{dick2021balancing}. Note that the immersive technologies essentially need the collection of the biometric data which acts as a baseline to create immersive experience, however, this ongoing feedback information (includes biometrics) creates novel issues for user privacy. Although the data collection is the core functions of immersive technologies which distinguish these technologies from other consumer devices and applications, however we argue that we must consider the scale and sensitivity of the data and should develop approaches to balance the secure use and the innovation of this technology~\cite{online4}.\par 
The urgency of this research is underscored by the widespread adoption of AR/VR technologies across various domains, including entertainment, healthcare, education, and industrial applications. As these technologies continue to permeate our daily lives, the collection and analysis of biometric signals have the potential to unlock unprecedented insights into user behavior and physiological responses~\cite{online5}. However, this newfound potential comes hand-in-hand with concerns surrounding user consent, data misuse, and the risk of biometric data breaches. More importantly, the tracking of data without user’s consent is illegal and violates human rights (individual’s rights to privacy)~\cite{tugtekin2023dark}. Australia’s esafety Commissioner raised the concern that immersive technologies can be used for harms cyberbullying, grooming children for online sexual abuse, and image-based abuse (sharing intimate content of someone without their consent, including sexual extortion). By Gartner’s report, it is predicted that by 2026, 25\% of people will spend at least one hour a day in the immersive technologies for work, shopping, education, social and/or entertainment~\footnote{https://www.gartner.com/en/newsroom/press-releases/2022-02-07-gartner-predicts-25-percent-of-people-will-spend-at-least-one-hour-per-day-in-the-metaverse-by-2026 assessed on 07/09/2023.}. However, to establish the sense of community that is central to its mass adoption; safety and ethical aspects (balancing data privacy issues) are significantly important now than ever~\cite{o2023privacy}. If the problem is not addressed timely will pose big risks to businesses, damage a brand’s reputation, impact economy, and user’s privacy, eventually a threat to national security and economy. We emphasize that this is a very novel research issue, and we are among the early ones to investigate this sensitive problem (i.e., \textit{to mitigate the sensitive and biomertic data leakage from the immersive technologies devices without the participant's consent.}) \par
Overall, the promise of immersive experiences is coupled with the risks of harvesting sensitive information, specific to the psychological and sensory interactivity of the participant. Further it could adversely impact the psychography of user, eventually violates data privacy and human rights laws~\cite{heller2020watching}. Motivated from this, our work in this paper dives deep into the critical realm of safeguarding user privacy within the ecosystems of immersive technologies. Specifically, our work addresses the formidable challenge of increasing the privacy of biometric data (or signals) collected through AR/VR devices. We propose a simple and effective methodology to extract biometric signals from raw data, devoid of any associated metadata, and subsequently, implement robust blocking mechanisms in order to avoid the leakage of biometric or sensitive data. The  overall goal is to empower users with greater control over their biometric data, ensuring that sensitive information remains confidential and secure.\par 
We reiterate that we propose a framework for the extraction of biometric signals from a spectrum of unconventional data sources received from the AR/XR devices. We aim to extract these biometric signals from diverse data modalities, such as audio signals and other non-standard sources, without reliance on accompanying metadata. This process intends to render the traceability of specific biometric data to individual users exceedingly challenging for unauthorized entities.\par
The contributions of this paper are as follows:
\begin{enumerate}
\item Particularly, we pointed out that the psychography data leakage issue is not well investigated in the state-of-the-art works. Most of the existing works focus on the design of biometric data-based authentication approaches to authenticate user, however, the issue of preventing biometric data leakage is largely ignored. We are the early ones to exploring and evaluating this critical research problem in immersive technologies context. This is our major contribution which aspires to make a substantial impact onto the ongoing discourse on biometric data privacy within AR/VR realms.   
\item By pioneering a simple and effective two-stage methodology for biometric signal extraction from unconventional data sources and fortifying their security, we aim to establish a foundation for a future where privacy and innovation coexist harmoniously in augmented and virtual reality. 
\item We propose a framework to mitigate sensitive data privacy leaks in immersive technology domain. The robust performance evaluation of the proposal, using six data sets, is conducted to show that the proposed solution is applicable, feasible and effective in this problem domain.  
\end{enumerate}
\par
The rest of the paper is organized as follows. We first review the related work in Section II. Then, in Section III, we present the proposed approach. In Section IV, we present the results of our experiments and some discussions on the performance of the proposed approach. Section V highlights the open issues within the scope of the problem context. Finally, in Section VI, we summarize the paper and provide some future work directions.

\begin{figure*}[t]

\centering

\includegraphics[width=5in, height=2in]{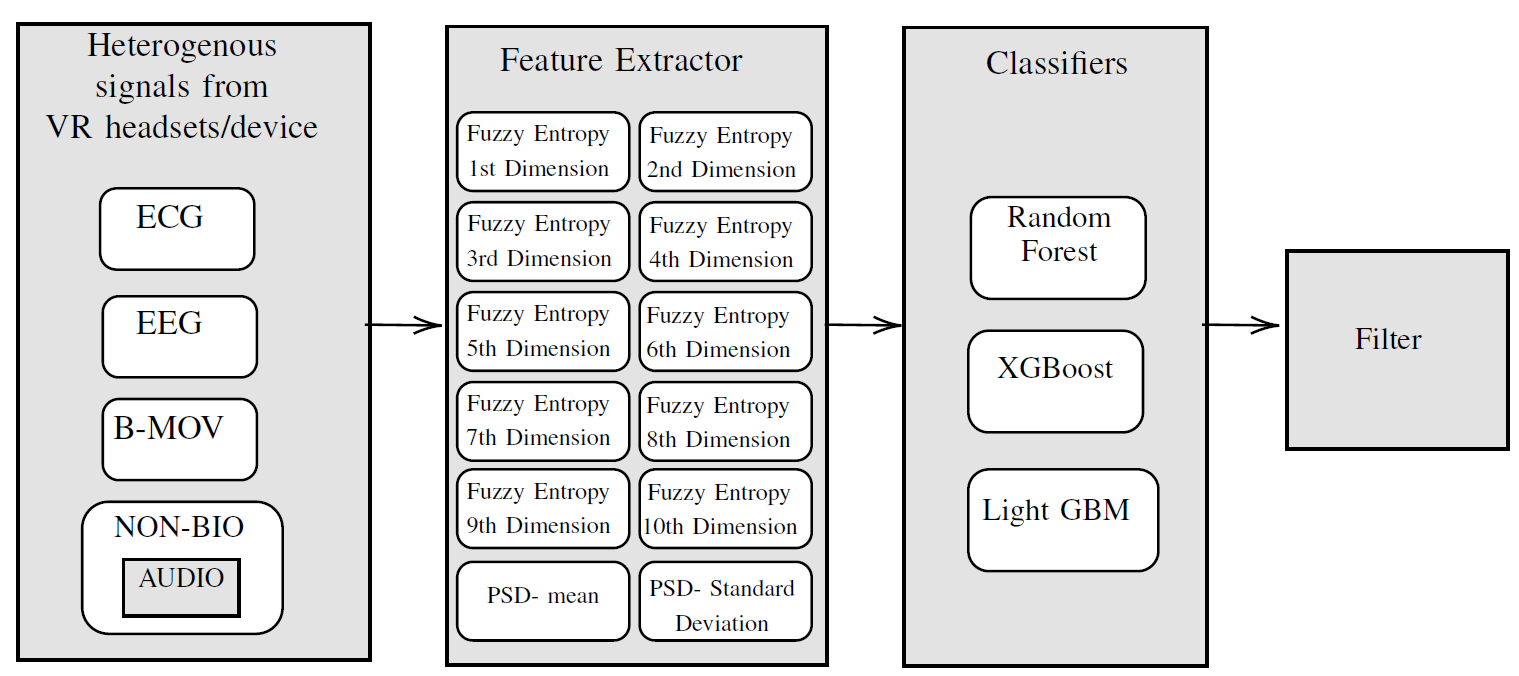}
    \caption{The high level view of the proposed approach. The approach comprises two primary modules: a feature extractor and classifiers. The feature extractor processes raw signal data, extracting a range of features. Subsequently, these extracted features serve as input for the classifiers, which employ algorithms like random forest, XGBoost, and LightBGM. The classifiers classify the signals or data into its correct category, and the output is routed through a filter. The filter selectively allows the passage of non-biometric signals, screening out any biometric signals.}
    \label{fig:architecture}
\end{figure*}

\section{Related Work}
In this section, we review the very recent literature related to avoid the biometric data leakage of participants from immersive technology devices~\cite{chen2023web3}.  \par
In Metaverse, \textit{``the sensory information and actuator-related information are exchanged between virtual and real worlds via IEEE 2888.1 and IEEE 2888.2 standards, respectively. Besides, the definition, synchronization, and mission control data are defined by the IEEE 2888.3 standard for digital things (i.e., virtual objects)"}~\cite{9880528}. Given these standards and the other enabling technologies researchers are using in Metaverse, we note that most of the existing works focus on some typical issues such as Identity Theft~\footnote{https://threatpost.com/nft-investors-lose-1-7m-in-opensea-phishing-attack/178558/, accessed on 06/11/2023}, Impersonation Attack, Avatar Authentication Issue,  Trusted and Interoperable Authentication,  Unauthorized Data Access, Misuse of User/Avatar Data, etc.~\cite{9880528, jiang2021reliable, han2022dynamic, lee2021creators, huynh2023artificial, falchuk2018social, ning2023survey}. To the best of our literature review, we note that very little has been explored to mitigate the issue of Biometrically inferred data~\cite{9880528, ning2023survey, huynh2023artificial, lee2021creators}. \par 
To better comprehend this, we further elaborate this issue. We understand that the, immersive technologies such as VR/XR or Metaverse rely on tracking, processing, adapting to the participant(s) sensory experience. Failure to accommodate for the expected visual, auditory, locomotive effect of the system disrupts the participant(s) vestibular system and causes simulator sickness~\cite{o2023privacy}. Thus, the hardware must monitor the user in deeply intimate and pervasive ways~\cite{heller2020watching}. Whilst immersive technologies rely on tracking and tracing of the participant(s) senses, to experience verisimilitude and presence, there is the substantial risk of private and sensitive biometric data being collected, sold and/or hacked/ and/or harvested by multiple third parties for further use, misuse, and/or abuse. 
Biometrically inferred data is a collection of datasets resulting from information inferred from behavioural, physical, and psychological biometric identification technique~\cite{heller2020reimagining}. \par
Interestingly, it is important to note that many definitions of biometrics rely on narrow physiological categories of data that may not cover data captured in immersive systems, and biometric identifiers and information is only covered if it is for authentication purposes~\cite{online3}. As we have mentioned above that very little has been explored or investigated in this direction, we are among the early ones sketching this problem and the state-of-the-art in detail. The authors~\cite{cheng2023towards} proposed zero-trust user authentication scheme fundamentally based on biometrics-based authentication that is suitable for continuously authenticating VR users. Further, in~\cite{10159439}  the authors design an avatar’s two-factor identity model to ensure the verifiability of avatar’s virtual identity and physical identity. They used biometric features in their design, which in our opinion is violation of the privacy of users, i.e, if the headset or device collects the biometric features without users consent is a clear violation of human rights or privacy of individual. Many other existing works such as~\cite{luo2021fa, bernadelli2021dynamic, chen2022query2set, zhang2020learning, zhang2018continuous, ryu2022design} proposed biometric-based authentication approaches to authenticate users, these approaches are divided into two sub-categories, physiological biometric-based and behavioral biometric-based.\par 
We extensively read the recent works~\cite{9880528, ning2023survey, huynh2023artificial, lee2021creators} and note that much of the existing works is around the policy design, proposing novel regulatory framework, etc.~\cite{shariff2023misogyny, allouzi2023adequate, wu2023financial, kostick2023ethical} but none has been explored significantly from the technical side of this problem.

\section{The Proposed Approach}

The high level view of the proposed framework is shown in Fig. 1. Note that the VR devices rely on numerous sensors to measure the motions made by the user/participant, for example infrared cameras to track eye movement, etc. Particularly, the Oculus Rift headset features an accelerometer, a gyroscope, and a magnetometer, all of them work together to track the rotation of the headset. The Oculus Rift also uses an infrared camera, which is used to track the position of the headset in relation to the room~\footnote{https://www.meta.com/en-gb/blog/quest/building-a-sensor-for-low-latency-vr/, accessed on 05/11/2023}. In the recent studies researchers have collected various signals from the VR devices such as Electroencephalography (EEG) signals, continuous blood pressure monitoring using low-cost motion sensors on AR/VR headsets, etc. We reiterate that there is clear threat of volition of user data privacy by capturing these biometric signals without user's consent. \par 
For example, eye tracking biometric data can cause pupil dilation in response to visual stimuli, brain activity related physiological data can cause increased levels of emotional activation, etc. The predictive behavioural analysis of biometric data, known as Biometric Psychography, is used to determine the emotional state of the participants upon viewing or interacting with products in immersive technologies. Psychography data feeds into misinformation, social, political, and electoral interference, etc.  Additionally, psychographic biometric data can be sold on the dark web and can pose threats to national security. Therefore, in our proposed framework we apply a simple two-stage approach (feature extraction and classification stage, filter stage, see Fig. 1)  to stop the leakage of biomertic data. Note that our aim is to prevent the leakage of biomertic data from VR devices.  \par
Firstly, as seen in Fig. 1, the framework comprises two primary modules or stages: stage 1 is a feature extractor and classifiers, and stage 2 is the filter. The feature extractor processes raw signal data, extracting a range of features. Subsequently, these extracted features serve as input for the classifiers which predict the signal’s category (biomertic or non-biomertic), and the output is routed through a filter. The filter selectively allows the passage of non-biometric signals, screening out any biometric signals. In our study, individual signals are collected separately from various datasets 
adjusting the number of samples for the varying sample rates of the datasets.\par 
The work-flow is presented in Algorithm 1. The proposed approach consists of three main components. Note that the proposed algorithms (or the overall approach) can be a part of the VR device's software. This can be a separate module or be an integral part of the device to mitigate the issue. In the following literature we elaborate the working of the three components of the proposal.    \par
\textcolor{black}{Note that all the  biometric signals (ECG, EEG, Body Movement) were obtained from the PhysioNet datatset repository~\footnote{https://physionet.org/about/database/}. The audio signals (GTZAN dataset) is claimed to be used widely for music genre classification task and we have used this audio corresponding to non-biometric signal. Our work is aligned with~\cite{melzi2022overview} and~\cite{bevsenic2022picking}}.

\textbf{\textit{Stage 1A:- Feature Extraction module}} - In the existing literature, various features have been systematically collected to characterize different types of signals. For instance, in the context of a universal and privacy-preserving EEG-based authentication system, proposed methods include autoregressive models, power spectral density analysis, and the utilization of wavelet transform for EEG signal~\cite{ Towards-privacy-EEG }. Similarly, in the case of stable EEG Biometrics using convolutional neural networks and functional connectivity, models incorporating power spectral density functions and fuzzy entropy measures are employed. Furthermore, affective EEG-Based person identification, utilizing deep learning techniques, relies on power spectral density estimation through Welch methods for individual identification based on EEG data~\cite{Affective-EEG-Based}. Entropy-based methodologies have gained substantial recognition in characterizing physiological signals, spanning EEG, ECG, and blood pressure recordings. Notably, various studies have applied fuzzy entropy metrics for signal analysis and comparison purposes~\cite{Fuzzy_Entropy_Metrics}. In addition, some research papers have incorporated distinctive features like PR and RR intervals for authentication objectives. In another context, studies has predominantly utilized spectral features, including spectral entropy, roll-off, and spread, alongside time-domain features such as energy metrics to authenticate individuals using head movements~\cite{head_authentication}. \par 
In our work, we adopted features that are common to ECG, EEG, and body movement signals, encompassing fuzzy entropy calculations spanning dimensions from 1 to 10, as well as statistical attributes of power spectral density such as mean and standard deviation. These feature selections are consistent with the existing works in immersive technologies context to authenticate a participant. Further the selection of these features were made with a focus on robustness and universality across signal types. The extracted features are passed onto the classifier module. \par    
\textbf{\textit{Stage 1B:- Classifier module}}-  The extracted features served as the basis for training various machine learning models, encompassing Random Forest~\cite{random-forest}, XGBoost~\cite{xg-boost}, and LightGBM~\cite{lightgbm}. Given the notable performance achieved by these individual models on the test data, there was no imperative need to pursue an ensemble of models. It is noteworthy that both XGBoost and LightGBM inherently exhibit ensemble characteristics, making the pursuit of additional ensemble strategies redundant in this particular context. To assess the robustness of the models under real-world conditions, various signal distortions were introduced. It is essential to add noise in the data as researchers have mentioned that the collected data can be very noisy due to various psychological effects~\cite{alain2023introduction}. In our work, the introduced distortions encompassed horizontal and vertical scaling, additive white Gaussian noise, and the use of smaller signal segments compared to the segments used for model training.Subsequently, the classification accuracy \textcolor{black}{and F1 score} was used  to evaluate the performance of these models.  \par
\textbf{\textit{Stage 2: Filter module}}- Signal passage is determined by the classifier's probability prediction. If the classifier identifies the signals as ECG, EEG, or body movement data, they are barred from progression. The work-flow of the classifier and the filter is presented under Algorithm 2. Note that the NON-BIO denotes the non-biomertic signal category. 

\begin{table}
	\label{algo:overall}
	\begin{tabular}{l}
		\midrule
		\textbf{Algorithm 1:} The Proposed algorithm for the overall approach\\
		\midrule
		\textbf{Input:}  \  {$s(t)$} (\textcolor{black}{Heterogeneous signals from VR headset/device})\\ \\

            $ class \leftarrow Classifier (s(t)) $ \\
            $ allowed\_to\_pass \leftarrow Filter(class) $ \\ \\
            
            \textbf{Outcome:} \textit{allowed\_to\_pass} \\

		\bottomrule
	\end{tabular}
\end{table}

\begin{table}
	\label{tab:freq}
	\begin{tabular}{l}
		\midrule
		\textbf{Algorithm 2:} The proposed approach of classifier algorithm\\
		\midrule
		\textbf{Input:}  \  \textit{$t^*$} (Minimum time length of a signal)\\
            \ \ \  \ \ \ \ \ \ \  \textit{$s(t)$} (Input signal)\\
            \ \ \  \ \ \ \ \ \ \  \textit{$ D_k $} (k = \{ \text{ECG, EEG, B-MOV,  \textcolor{black}{NON-BIO}} \})\\
            \ \ \  \ \ \ \ \ \ \  \textit{$g_k(t)$} (Signals for $t \geq t^*$, where $k \in D_k$) \\
            \ \ \  \ \ \ \ \ \ \  $\exists f: g(t) \rightarrow \theta_G$ (feature extractor function) \\ \\

            \textbf{Variables:}  \ \  \textit{$max\_probability=0$} \\
            \ \ \  \ \ \ \ \ \ \ \ \ \ \ \ \  \textit{class} \\ \\

            \textbf{if} $t >t^* $: \\
   
            \ \ \ $ \theta_S \leftarrow f(s(t))$ \\
            \ \ \ $ \theta_G \leftarrow f(g(t))$ \\
            \ \ \ \textbf{for} $k \in D_k $: \textbf{do} \\
		\ \ \ \ \ \ \ \ \ \  \textbf{if} $Pr( \theta_S| \theta_G) > $ \text{max\_prob}: \\
            \ \ \ \ \ \ \ \ \ \ \ \ \ $\text{max\_probability} \leftarrow Pr(S|G_k)$ \\
            \ \ \ \ \ \ \ \ \ \ \ \ \ $\text{class} \leftarrow  k $ \\
            \ \ \ \textbf{end for} \\
            \textbf{end if} \\ \\
           \textbf{if} $class ==  \textcolor{black}{NON-BIO}  $: \\
              \ \ \ \ \ \ \ \ \ \ allowed\_to\_pass = \textbf{TRUE} \\
            \textbf{else} \\
              \ \ \ \ \ \ \ \ \ \ allowed\_to\_pass = \textbf{FALSE} \\
            \textbf{end if} \\          
            \\
            \textbf{Outcome:} \textit{allowed\_to\_pass} \\

		\bottomrule
	\end{tabular}\
\label{algo:classifier}
\end{table}

\section{Performance Evaluation}
In this section, we have evaluated the performance of our scheme in the detection of signals into ECG, EEG, Body Movement, and NON-BIO class. We have firstly presented the setups of our experiments and discussed the dataset we have used in our work. Following this, we have presented the results of our experiments.\par
We employ open-source tools and the real datasets to show the Proof of Concept (PoC) and the effectiveness of the proposed framework. We use six datasets comprising biometric signals, including 1) A large scale 12-lead electrocardiogram database for arrhythmia study, 2) Apnea-ECG Database, 3) Auditory evoked potential EEG-Biometric dataset, 4) Body Sway When Standing and Listening to Music Modified to Reinforce Virtual Reality Environment Motion, 5) GTZAN Dataset for Music Genre Classification and 6) Norwegian Endurance Athlete ECG Database, consisted of multiple channels of data for each individual. These multichannel data were systematically segmented into non-overlapping segments of 8 seconds each, a process employed for training and analysis purposes. Below we provide the detailed description of each dataset we used in our work. 
\subsection{Data-set description}
\textbf{\textit{1. Large Scale 12-Lead Electrocardiogram Database for Arrhythmia Study}}~\cite{ECG_12_electrode}- This extensive research database offers a wealth of 12-lead electrocardiogram (ECG) signals sampled at a rate of 500Hz. With a primary goal of advancing studies on arrhythmia and cardiovascular conditions, it comprises data from 45,152 patients. Collaboratively created by Chapman University, Shaoxing People’s Hospital, and Ningbo First Hospital, the dataset focuses on atrial fibrillation, a cardiac condition of significant public health concern. It encompasses various common rhythms and cardiovascular conditions, meticulously labeled by expert professionals.\par

\textbf{\textit{2. Auditory Evoked Potential EEG-Biometric Dataset}}~\cite{Auditory_Evoked-EEG}-This dataset, recorded at a sampling rate of 200Hz, includes over 240 two-minute EEG recordings from 20 volunteers. It encompasses resting-state EEG signals with both eyes open and eyes closed, as well as experiments involving auditory stimuli. The primary objective is to facilitate the development of an EEG-based biometric system. EEG signals, which reflect brain activity, have diverse applications, including biometric authentication. This dataset serves for a range of purposes, from simple analysis to comparisons between resting states with varying eye conditions and studying the effects of auditory stimuli. During data collection, subjects underwent EEG recording sessions with specific electrode placements according to the 10/10 international EEG system, participating in experiments that included resting-state conditions, auditory stimuli with different songs, and sessions with noise isolation. \par 
In our experiment, EEG recordings were selected from individuals as they listened to songs through in-ear headphones, as this scenario closely relates to VR/AR experiences. \textcolor{black}{EEG recordings obtained during individuals' auditory experiences through in-ear headphones were employed for the training of the classifiers. Additionally, the EEG recordings during the auditory experiences through bone-conducting earphones were subjected to various intentional distortions. These perturbed signals were subsequently utilized to evaluate the classifiers' resilience and performance under adverse conditions. }\par 
\textbf{\textit{3. Apnea-ECG Database}}~\cite{apnea}-This database comprises 70 records, divided into a learning set of 35 records and a test set of 35 records. The recordings, sampled at a rate of 100Hz, vary in length, but non-overlapping segments of 8 seconds were created from them. Each record includes a continuous digitized ECG signal, apnea annotations, and machine-generated QRS annotations.\par 

\textbf{\textit{4. Body Sway When Standing and Listening to Music Modified to Reinforce Virtual Reality Environment Motion}}~\cite{body_sway}- This dataset, recorded at a sampling rate of 1000Hz, was meticulously collected to investigate the influence of music integrated into a VR environment on body sway. It comprises measurements of body sway obtained from subjects standing on a balance platform under various visual and auditory conditions within the VR setup. The dataset's relevance lies in its exploration of the impact of music on body sway, particularly within the context of a VR environment. Increased body sway is often associated with falls, especially in aging and patient populations. Therefore, this dataset provides valuable insights into the complex interplay between sensory inputs and postural stability during VR experiences. Researchers can utilize the high-resolution data to gain a deeper understanding of how music affects body sway in VR scenarios and its potential implications for various applications.\par 
\textbf{\textit{5. GTZAN Dataset for Music Genre Classification}}~\cite{gtzan}- 
It stands as a renowned resource in the field of music genre classification, celebrated for its diverse assortment of audio clips representing a wide spectrum of music genres, encompassing rock, jazz, pop, classical, blues, and more. Each audio clip, typically spanning a few seconds, is meticulously labeled with its corresponding genre. Researchers and practitioners frequently turn to this dataset for developing and assessing machine learning models and algorithms designed to automatically categorize music into various genres based on audio content. Notably, the original audio clips within the GTZAN Dataset are sampled at a standardized rate of 44.1kHz. However, to enhance computational efficiency, we  have opted to down-sample the audio data to 1000Hz, striking a balance between maintaining audio quality and optimizing processing speed.  \textcolor{black}{This dataset has been used corresponding to the `NON-BIO' class. This helped us to evaluate the ability of the model to differentiate ECG, EEG, and Body movement signals apart from the non-bimoetric signals.}  \par 
\textbf{\textit{6. Norwegian Endurance Athlete ECG Database}}~\cite{norwegian_ecg}-
This dataset features 12-lead ECG recordings obtained from a cohort of 28 elite athletes across diverse sporting disciplines in Norway, offers meticulously recorded data sets with an impressive standard of precision. These ECG recordings, acquired at a consistent and typically high sample frequency of 500Hz, ensure a comprehensive and finely detailed portrayal of the athletes' cardiac electrical activity, especially during periods of resting conditions. Moreover, this dataset has been instrumental in rigorously evaluating the robustness and reliability of predictive models when subjected to various forms of data distortion and alteration, advancing the field's understanding of model performance under challenging conditions.

\begin{figure*}
\centering
    \includegraphics[width=7.2in, height=1.8in]{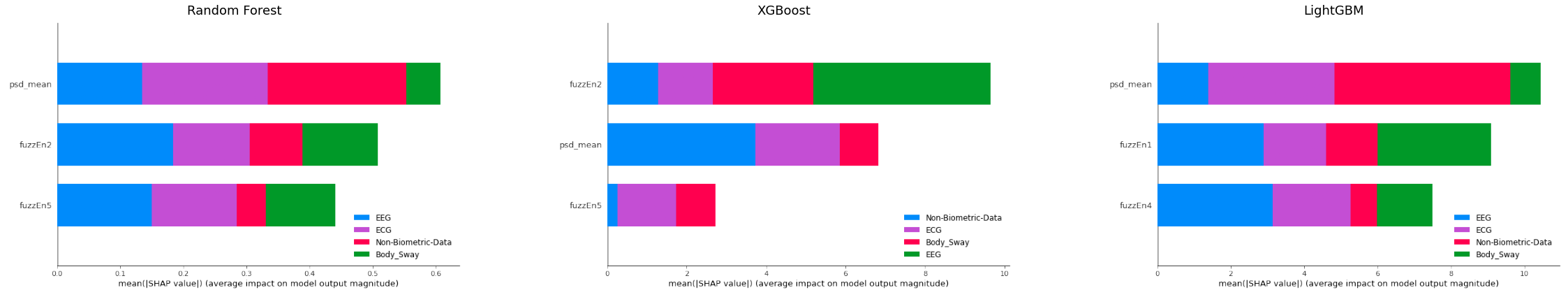} 
\captionsetup{labelfont = bf, font = footnotesize}
\caption {\textcolor{black}{Feature importance was compared across Random Forest, XGBoost, and LightGBM classifiers using the {MultiModalBioAudio dataset}, SHAP values were computed following the execution of a forward feature selection algorithm aimed at identifying the most influential features for each model. The analysis focused exclusively on the impact of features selected through the forward feature selection process.}}
    \label{fig:shap}
\end{figure*}

\subsection{Experimental Setup}
The experiments were carried out on a computing system having $64$-bit operating system with an x$64$-based processor, $8.00$GB of RAM ($7.71$GB usable), and Intel(R) Core(TM) i$7-11800$H processor running at a base clock speed of $2.30$GHz, all under normal operational conditions. We initiated our research by first developing a binary classifier to distinguish between ECG~{\cite{ECG_12_electrode}} and non-ECG signals~\cite{Auditory_Evoked-EEG}. Following the successful implementation of this binary classification model, we expanded our investigation to create a multi-class classifier capable of categorizing various biometric signals, including ECG~{\cite{ECG_12_electrode}}~{\cite{apnea}}, EEG~{\cite{Auditory_Evoked-EEG}}, body movements~{\cite{body_sway}}, and audio signals~{\cite{gtzan}}. \par

In our study, a comprehensive dataset comprising 2000 ECG signals was utilized, with each group of 500 ECG signals drawn from various sources, including the APNEA dataset~{\cite{apnea}}, the APNEA dataset with the introduction of AWGN noise, the Arrhythmia Study dataset, and the Arrhythmia Study dataset~{\cite{ECG_12_electrode}} with the same level of added noise. Additionally, we incorporated 1,500 EEG signals~{\cite{Auditory_Evoked-EEG}} and 750 signals sourced from the Body Sway dataset~{\cite{body_sway}}, all of which were complemented by an additional 1,000 audio signals~{\cite{gtzan}} for non-biometric data. Henceforth, this dataset, in our work, is denoted as the ``MultiModalBioAudio Dataset" for academic reference. To ensure a rigorous evaluation, we partitioned the entire dataset into a training set comprising $70\%$ of the data and a separate test set comprising the remaining $30\%$. This meticulous dataset composition allowed us to comprehensively assess the performance of our models across various signal types, thereby enhancing the robustness and validity of our findings.\par
Following the training of our models, we conducted rigorous testing to evaluate the robustness of our models across various signal types. Specifically, the study focused on signal sources from the Norwegian Endurance Athlete ECG Database and the Auditory Evoked Potential EEG-Biometric Dataset. The latter dataset comprised EEG recordings from individuals who were exposed to auditory stimuli through bone-conducting earphones. The recordings from both datasets were subject to various forms of distortions in our analysis. These introduced distortions encompassed the addition of noise with varying standard deviations, as well as vertical and horizontal scaling with distinct scaling factors. It is important to highlight that no preprocessing steps were applied to the experimental setup, ensuring a comprehensive assessment of model performance under these varied and challenging conditions. This thorough testing allowed us to ascertain the models' resilience and adaptability across a spectrum of real-world scenarios and signal alterations. The equations associated with different distortions are given in Table~\ref{table:distortions}
\begin{table}[ht]
\centering
\captionsetup{labelfont = bf, font = footnotesize}
    \caption{Equations for different distortions. {Here z(t), s(t), and m(t) represents the distorted signal and raw signal, respectively.}} 
    \label{table:distortions}
    \captionsetup{labelfont = bf, font = footnotesize}
    \begin{tabular}{c c}
    \hline
    Distortion Type & Equation \\
    \hline
    Horizontal Scaling & $z(t) = s(  \alpha \cdot t)$ \\
    Vertical Scaling & $z(t) = \alpha \cdot s(t)$ \\
    Additive White Gaussian Noise & $z(t) = s(t) + \textcolor{black}{\mathcal{N}(x; 0, \sigma)}$ \\
    Smaller Signal Segments & $ Algorithm $3 $ $ \\
    \hline
    \end{tabular}
\end{table}

\begin{figure*}
    \includegraphics[width=7.2in, height=3in]{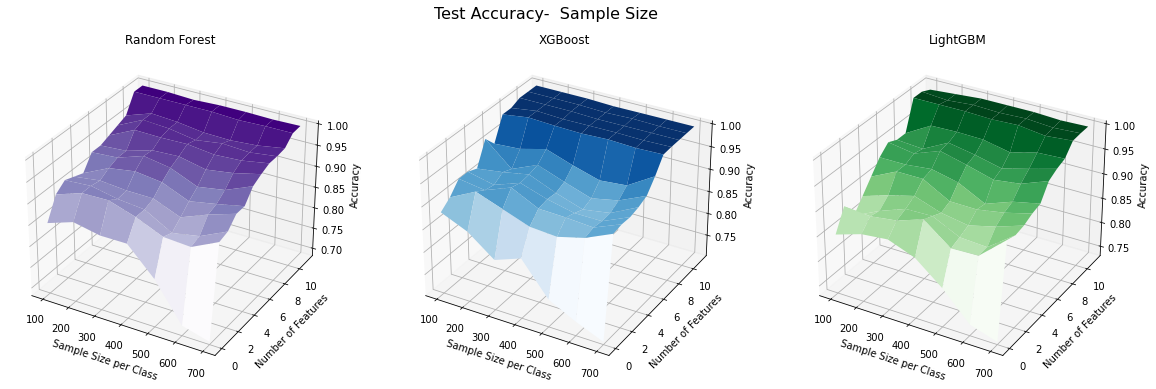}
    \captionsetup{labelfont = bf, font = footnotesize}
    \caption{\textcolor{black}{Accuracy of the Random Forest Classifier, XGBoost Classifier and LightGBM Classifier for different features and
sample size.}}
    \label{fig:Test-accuracy}
\end{figure*}


\begin{table}
	\label{algo:cropping}
	\begin{tabular}{l}
		\midrule
		\textbf{Algorithm 3:} The Random Cropping Algorithm\\
		\midrule
		\textbf{Input:}  \  \textit s(t)\\

            \ \ \  \ \ \ \ \ \ \  \textit {T} (Length of the output signal) \\
            \textit{start\_time } $\leftarrow$  Random float $x$ ; $x \in (0,t-T) $ \\
           ${z(t)=s(t^*)}$ for $ t^* \in [start\_time, start\_time+ T] $ \\

            \textbf{Outcome:} \textit{$z(t)$} \\

		\bottomrule
	\end{tabular}
\label{algo:filter}
\end{table}

\subsection{Results}
In Table~\ref{tab:training-testing-times}, we present the training and testing times for our machine learning models. Notably, the models exhibit remarkable efficiency in both training and testing phases. These precise timings provide a comprehensive understanding of the models' computational performance on the specified hardware and software configuration, offering valuable insights for further analysis and optimization.
\begin{table}[ht]
\centering
\captionsetup{labelfont = bf, font = footnotesize}
\caption{Training and Testing Times {of the MultiModalBioAudio Dataset for the three classification algorithms. }}
\label{tab:training-testing-times}
\begin{tabular}{lcc}
\hline
\textbf{Model} & \textbf{Training Time (s)} & \textbf{Testing Time (s)} \\
\hline
Random Forest & 0.291 & 0.014 \\
XGBoost & 0.215 & 0.003 \\
LightGBM & 0.224 & 0.006 \\
\hline
\end{tabular}
\end{table}

\par
Also, in our experimentation, we first evaluate the critical features for each classifier, extracting valuable insights into their significance. \textcolor{black}{We also ran forward feature selection algorithm to find the most effective combination of the extracted features. We used SHAP~\cite{SHAP} to gain insights on how the selected features contributed to the prediction of each model} These feature importance rankings and their contribution are visually presented in Figure~\ref{fig:shap}. \textcolor{black}{The figure suggests that the Random Forest model and LightGBM places considerable importance on the mean of power spectral density, whereas for XGBoost, the 2nd dimension of fuzzy entropy emerges as a significant contributor. Additionally, it is noteworthy that the power spectral density distribution of the signals plays a crucial role in the predictions of all models.}  Following this, we proceeded to train the classifiers and subsequently examined their performance by varying the sizes of the training datasets. \textcolor{black}{The individual features were sorted in the ascending order based on the importance score provided by each classifier. }\par
During this process, we incrementally added features to the models, starting from the least important and progressing to the most important ones. The experimental results, which showcase the variations in classification accuracy with changes in training data sizes and feature incorporation, can be observed in Figure~\ref{fig:Test-accuracy}. The results from the figure illustrate that using large number of samples for one feature results in overfitting of all the models, and when the samples and features increase, there is a consistent increase in the classification accuracy. However, when the number of features {for the MultiModalBioAudio Dataset }increase to more than 10, the number of sample size for training  the test accuracy is saturated near 99\% irrespective of the number of training samples for each class. \par
Note that Biometric signals obtained from AR/VR headsets can be subject to both unintentional and intentional modifications. Unintentional alterations typically result from natural noise interference. In contrast, intentional modifications may be introduced by adversaries aiming to evade our proposed security measures and gain unauthorized access to the biometric data. These intentional modifications can encompass actions such as horizontal and vertical scaling, as well as random segment alterations applied to the output of the AR/VR headset. The relevant results are discussed below.\par
 \begin{figure*}[t]
    \centering
     \captionsetup{labelfont = bf, font = footnotesize}
    \subfloat[]{\includegraphics[width=3.7in, height=5in]{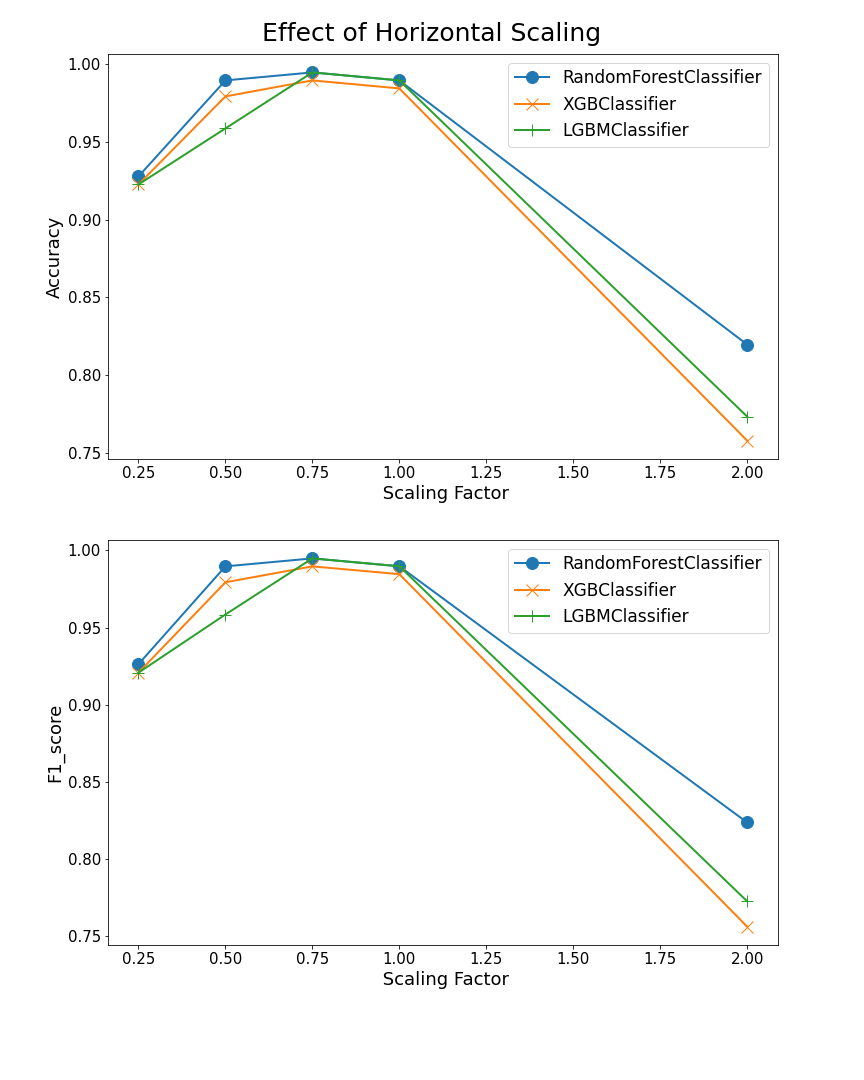}}
    \subfloat[]{\includegraphics[width=3.7in, height=5in]{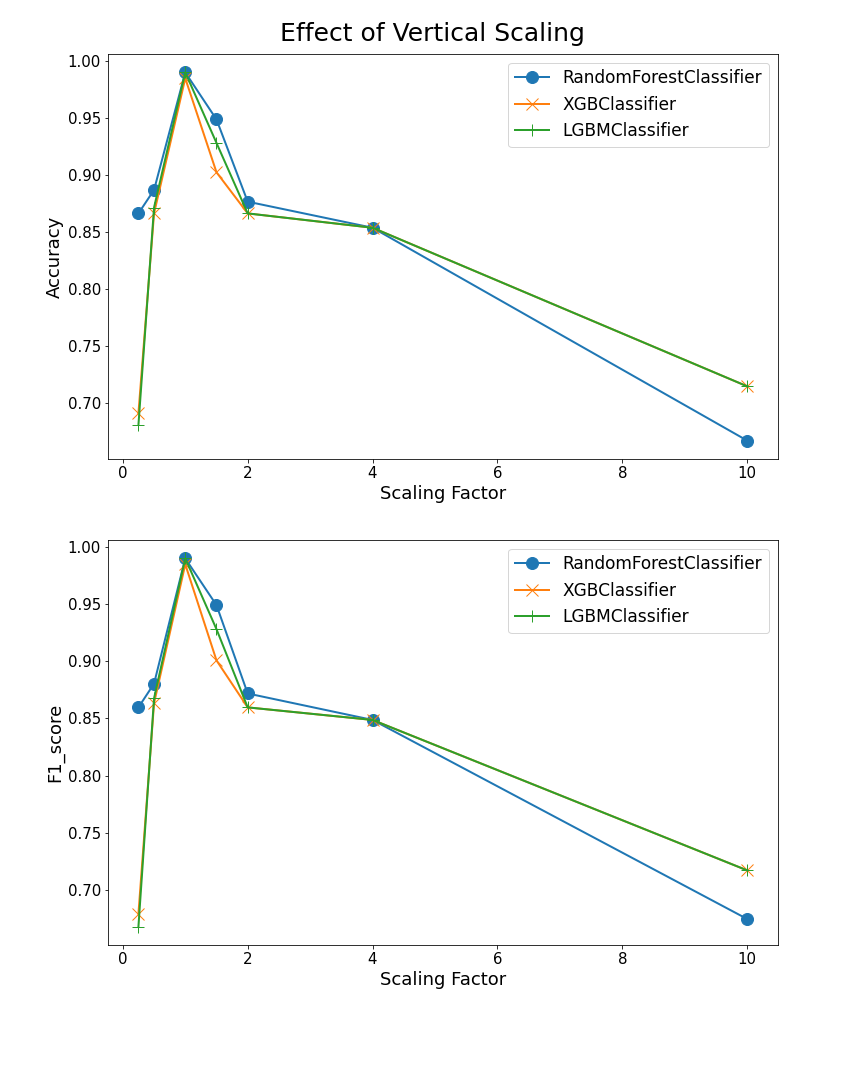}}
    \caption{Accuracy \textcolor{black}{and F1 score} of the Random Forest Classifier, XGBoost Classifier, and LightGBM Classifier for different factors of a) horizontal scaling and b) vertical scaling. \textcolor{black}{The signals were horizontally and vertically scaled corresponding (x-axis) to the equations $z(t) = s(\alpha· t)$ and $z(t) = \alpha · s(t)$ where z(t) is the scaled signal $\alpha$ is the scaling factor and s(t) is the original signal.}}
   \label{fig: horizontal-scaling}
\end{figure*}
We conducted an extensive assessment of the classifiers trained using all twelve features. This comprehensive evaluation involved measuring their classification accuracy on a dataset consisting of  \textcolor{black}{ 50 signals for ECG, EEG, body movement, and audio, each subjected to different types of distortions. The equations corresponding to various types of distortions are presented in Table~\ref{table:distortions}. Vertical scaling distortion involves multiplying the original signals by a factor $\alpha$, while horizontal scaling entails re-sampling the signal with different time scaling factors. Noise distortion is introduced by adding samples from a normal (Gaussian) distribution with a zero mean and various standard deviations. Additionally, cropping distortion involves randomly cropping the original signals at different time points to create windows corresponding to various time segments.}The findings were notably intriguing: signals that underwent distortions such as horizontal and vertical scaling, Fig.~\ref{fig: horizontal-scaling} a and b, respectively.\par 
\textcolor{black}{It is seen in the Fig.~\ref{fig: horizontal-scaling} (a) that till scaling factor is 1, i.e., $\alpha= 1$, the accuracy and the F1-score results are showing consistent results. After $\alpha=1$ the accuracy and F1 score drops significantly. Note that we want the proposed model to provide high accuracy and better F1 results even at very high values of $\alpha$. This is to validate that how reliable the the proposed solution is in the presence of adversarial activity, i.e., an adversarial attempt to temper the data. Therefore finding the optimum value of the scaling factor is a matter of further investigation of finding a trade-off between the utility and the privacy of the model which we aim to tackle entirely a new paper. Similarly in Fig.~\ref{fig: horizontal-scaling} (b), we note the same pattern of results.}\par 

\begin{figure}
\centering
     \includegraphics[width=3.7in, height=5in]{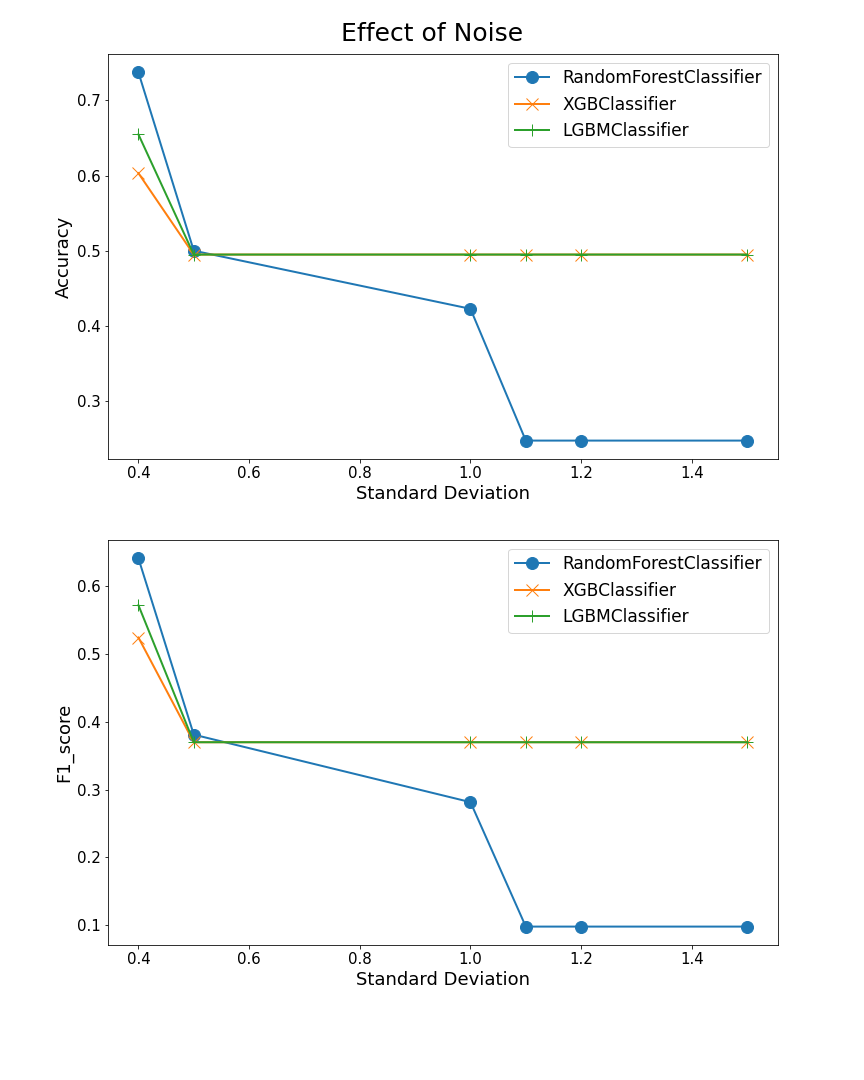}
    \captionsetup{labelfont = bf, font = footnotesize}
    \caption{\textcolor{black}{Accuracy \textcolor{black}{ and F1 score } of the Random Forest Classifier, XGBoost Classifier and LightGBM Classifier for different standard deviation values for AWGN. It is worth noting that the LightGBM classifier and XGBoost classifier exhibit similar behavior. \textcolor{black}{Noise was added to the original signal corresponding to the equation $z(t) = s(t) +\mathcal{N} (x; 0, \sigma)$ where z(t) is the output signal, s(t) is the original signal and $\mathcal{N}$ represents the normal distribution.}}}
    \label{fig:noise}
\end{figure}

\begin{figure}
\centering
   \includegraphics[width=3.7in, height=5in]{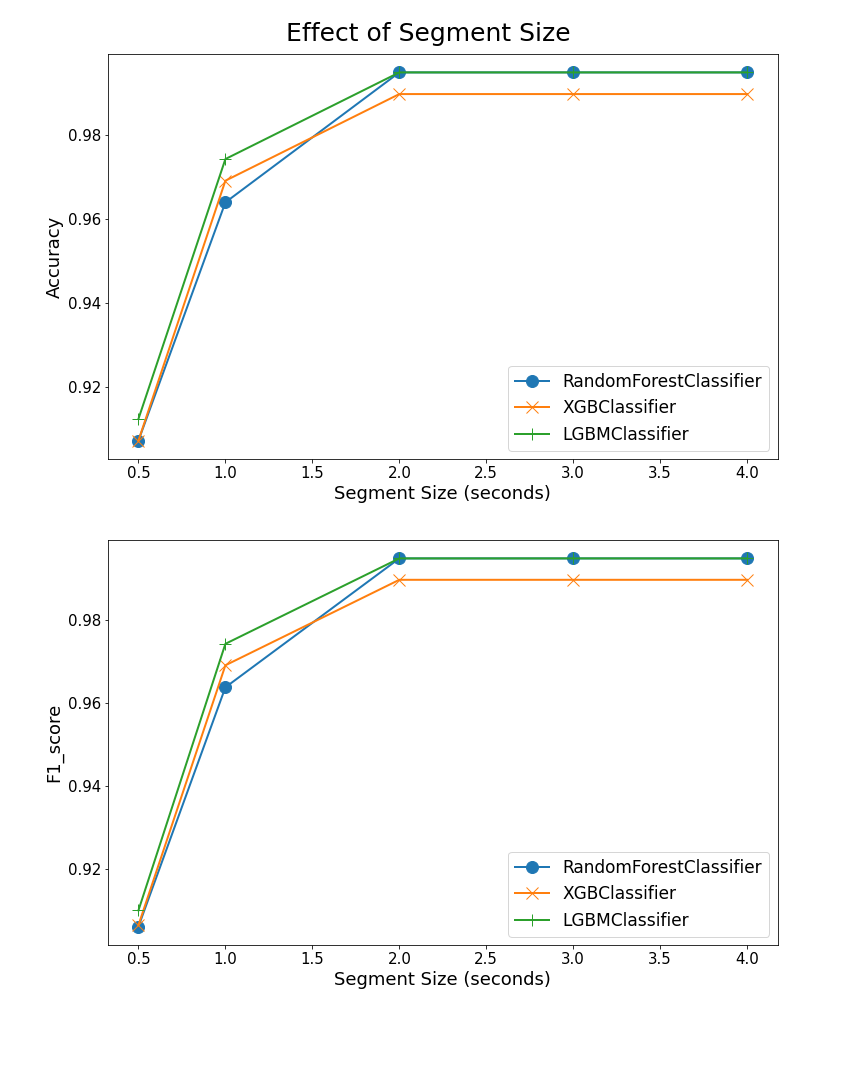}
    \captionsetup{labelfont = bf, font = footnotesize}
    \caption{\textcolor{black}{Accuracy \textcolor{black}{and F1 score} of the Random Forest Classifier, XGBoost Classifier, and LightGBM Classifier for different time lengths of signals.}}
    \label{fig:segemnt-size}
\end{figure}

Further, the results related to noise injection and  random segment alteration are shown in Fig.~\ref{fig:noise}, and  Fig.~\ref{fig:segemnt-size}, respectively.  \textcolor{black}{ These are providing new insights into the effects of distortion. The overall performance of all classifiers declined when the signals underwent distortions. \textcolor{black}{As evident from Figure~\ref{fig:segemnt-size}, the }\textcolor{black}{segmentation results indicated that despite training the signals with 8 seconds of data, a mere 2 seconds proved sufficient for high-accuracy and F1 score classification}. Additionally, the random forest classifier exhibited the highest resilience to horizontal scaling, but performance deterioration occurred when the horizontal scaling factor surpassed 1 for all classifiers.} For a detailed breakdown of classification accuracy concerning each type of distortion, please refer to Figures~\ref{fig: horizontal-scaling},~\ref{fig:noise},and~\ref{fig:segemnt-size}.

\section{Further discussion and Open Issues}
Uniquely, we consider the interconnected and interdependent relationship between biometrics and human rights with national and international regulations, for example Consumer Data Rights in Australia (CDR) and the EU’s General Data Protection Regulation (GDPR). We develop a framework to ensure that a collective and enforceable decision can be made by governments across international jurisdictional boundaries to develop socially responsible avatars that strengthen our user and data privacy. The outcome of our work will shape the global trends of applying secure techniques (such as the one we propose) to be embedded or integrated in VR devices or the immersive technologies devices within the scope of technical standards and ethical frameworks. The direct beneficiaries will be the public and private cyber security organisations, human rights organisations, and the immersive technology industries as well as the participants of immersive technology. We accept that the work proposed in this paper is limited but noteworthy to a great extend that we are the early ones unearthed this ignored issue. The presented work is an early evaluation and there are further many open issues to address. More analysis needed on how the approach would impact immersive user experience and device functionality.\par 
\textit{The devices' Usability:-} Firstly, it is essential to assess the proposed approach's usability, for that further deep evaluation from its deployment context is needed to evaluate. It is true that the human condition is extremely variable and hence the design of any approach would always leave novel anomalies which eventually may exacerbate and potentially create a new form of discrimination. Also, with the scale of participants (humans) and the diversity of human conditions (variety and uniqueness) would further create complexity of biometric data analysis in regards label and then classify the data. So, in future, we aim to update the firmware of VR devices with the proposed approach's and then further investigate the impact of this on users immersive experiences and the behaviour/functionality of the device. This will give us further insights and directions to further explore this area.  \par 
 \textit{Informed Consent:-}  Secondly, the problem of biometric data privacy is linked to human rights as well. From technical perspectives, there will be complexities in creating the baseline of how to classify the outlier data. There will be more struggle with the data analysis part in such ways, therefore the alternative is to seek help from our legislation or to re-ignite the concept of informed consent in this application domain. We note that the existing legislation or standards (IEEE 2888.1, IEEE 2888.2, and IEEE 2888.3) do not adequately specify how biometric data is collected and stored in the context of immersive technologies. Therefore, improved regulation is necessary to help preserve data privacy and human rights as well as to strengthen national security. \par
Further research from policy frameworks and regulatory aspects will help balancing the opportunities of immersive technologies to help enable cross jurisdictional actionable decision-making in the immersive technologies whilst upholding the tenets of fundamental human rights for handling of personal biometric data. Essentially this will help to maintain the data privacy in trans-boarder data flow scenarios. Policymakers can create an innovation-friendly regulatory environment in immersive technologies in order to maintain user privacy by clarifying, updating, and harmonizing existing rules and introducing comprehensive privacy legislation.
\section{Summary and Future Work}
In the ever-evolving landscape of Augmented Reality (AR) and Virtual Reality (VR) technologies, the collection and utilization of biometric signals have emerged as pivotal components, offering immersive experiences and personalized interactions. Biometric signals, derived from physiological and behavioral characteristics, have the potential to enhance user experiences by adapting content and interfaces in real-time. However, the integration of biometric data into AR and VR environments raises significant concerns regarding individual privacy and data security. In this work, we pointed out that the collection of biometric data without user consent is a clear violation of privacy and human rights. To mitigate this issue, we proposed a simple and effective two-stage approach to defend user's privacy. Our robust approach of testing our proposal shows the effectiveness of the solution. Finally, in the further discussion section, we have provided some future research directions in this context.

\bibliographystyle{IEEEtran}
\bibliography{subsections/references.bib}

\begin{thebibliography}{10}
\providecommand{\url}[1]{#1}
\csname url@samestyle\endcsname
\providecommand{\newblock}{\relax}
\providecommand{\bibinfo}[2]{#2}
\providecommand{\BIBentrySTDinterwordspacing}{\spaceskip=0pt\relax}
\providecommand{\BIBentryALTinterwordstretchfactor}{4}
\providecommand{\BIBentryALTinterwordspacing}{\spaceskip=\fontdimen2\font plus
\BIBentryALTinterwordstretchfactor\fontdimen3\font minus \fontdimen4\font\relax}
\providecommand{\BIBforeignlanguage}[2]{{%
\expandafter\ifx\csname l@#1\endcsname\relax
\typeout{** WARNING: IEEEtran.bst: No hyphenation pattern has been}%
\typeout{** loaded for the language `#1'. Using the pattern for}%
\typeout{** the default language instead.}%
\else
\language=\csname l@#1\endcsname
\fi
#2}}
\providecommand{\BIBdecl}{\relax}
\BIBdecl

\bibitem{cipresso2018past}
P.~Cipresso, I.~A.~C. Giglioli, M.~A. Raya, and G.~Riva, ``The past, present, and future of virtual and augmented reality research: a network and cluster analysis of the literature,'' \emph{Frontiers in psychology}, p. 2086, 2018.

\bibitem{dhar2021augmented}
P.~Dhar, T.~Rocks, R.~M. Samarasinghe, G.~Stephenson, and C.~Smith, ``Augmented reality in medical education: students’ experiences and learning outcomes,'' \emph{Medical education online}, vol.~26, no.~1, p. 1953953, 2021.

\bibitem{10271790}
Z.~Tian, C.~Zhang, K.~Sood, and S.~Yu, ``Inferring private data from ai models in metaverse through black-box model inversion attacks,'' in \emph{2023 IEEE International Conference on Metaverse Computing, Networking and Applications (MetaCom)}, 2023, pp. 49--56.

\bibitem{shi2021face}
C.~Shi, X.~Xu, T.~Zhang, P.~Walker, Y.~Wu, J.~Liu, N.~Saxena, Y.~Chen, and J.~Yu, ``Face-mic: inferring live speech and speaker identity via subtle facial dynamics captured by ar/vr motion sensors,'' in \emph{Proceedings of the 27th Annual International Conference on Mobile Computing and Networking}, 2021, pp. 478--490.

\bibitem{8919581}
S.~Li, S.~Savaliya, L.~Marino, A.~M. Leider, and C.~C. Tappert, ``Brain signal authentication for human-computer interaction in virtual reality,'' in \emph{2019 IEEE International Conference on Computational Science and Engineering (CSE) and IEEE International Conference on Embedded and Ubiquitous Computing (EUC)}, 2019, pp. 115--120.

\bibitem{john2020security}
B.~John, S.~J{\"o}rg, S.~Koppal, and E.~Jain, ``The security-utility trade-off for iris authentication and eye animation for social virtual avatars,'' \emph{IEEE transactions on visualization and computer graphics}, vol.~26, no.~5, pp. 1880--1890, 2020.

\bibitem{dick2021balancing}
E.~Dick, ``Balancing user privacy and innovation in augmented and virtual reality,'' Information Technology and Innovation Foundation, Tech. Rep., 2021.

\bibitem{online4}
\BIBentryALTinterwordspacing
What are the biggest issues facing the metaverse? [Online]. Available: \url{https://www.xrtoday.com/virtual-reality/what-are-the-biggest-issues-facing-the-metaverse/#:~:text=Data%20protection%20and%20privacy%20are,that%20could%20violate%20their%20privacy.}
\BIBentrySTDinterwordspacing

\bibitem{online5}
\BIBentryALTinterwordspacing
Social internet is dead. get over it. [Online]. Available: \url{https://om.co/2023/10/15/social-internet-is-dead-get-used-to-it/}
\BIBentrySTDinterwordspacing

\bibitem{tugtekin2023dark}
U.~Tugtekin, ``The dark side of metaverse learning environments: Potential threats and risk factors,'' in \emph{Shaping the Future of Online Learning: Education in the Metaverse}.\hskip 1em plus 0.5em minus 0.4em\relax IGI Global, 2023, pp. 57--67.

\bibitem{o2023privacy}
J.~O'Hagan, P.~Saeghe, J.~Gugenheimer, D.~Medeiros, K.~Marky, M.~Khamis, and M.~McGill, ``Privacy-enhancing technology and everyday augmented reality: Understanding bystanders' varying needs for awareness and consent,'' \emph{Proceedings of the ACM on Interactive, Mobile, Wearable and Ubiquitous Technologies}, vol.~6, no.~4, pp. 1--35, 2023.

\bibitem{heller2020watching}
B.~Heller, ``Watching androids dream of electric sheep: immersive technology, biometric psychography, and the law,'' \emph{Vand. J. Ent. \& Tech. L.}, vol.~23, p.~1, 2020.

\bibitem{chen2023web3}
H.~Chen, H.~Duan, M.~Abdallah, Y.~Zhu, Y.~Wen, A.~El~Saddik, and W.~Cai, ``Web3 metaverse: State-of-the-art and vision,'' \emph{ACM Transactions on Multimedia Computing, Communications and Applications}, 2023.

\bibitem{9880528}
Y.~Wang, Z.~Su, N.~Zhang, R.~Xing, D.~Liu, T.~H. Luan, and X.~Shen, ``A survey on metaverse: Fundamentals, security, and privacy,'' \emph{IEEE Communications Surveys \& Tutorials}, vol.~25, no.~1, pp. 319--352, 2023.

\bibitem{jiang2021reliable}
Y.~Jiang, J.~Kang, D.~Niyato, X.~Ge, Z.~Xiong, and C.~Miao, ``Reliable coded distributed computing for metaverse services: Coalition formation and incentive mechanism design,'' \emph{arXiv preprint arXiv:2111.10548}, 2021.

\bibitem{han2022dynamic}
Y.~Han, D.~Niyato, C.~Leung, D.~I. Kim, K.~Zhu, S.~Feng, X.~Shen, and C.~Miao, ``A dynamic hierarchical framework for iot-assisted digital twin synchronization in the metaverse,'' \emph{IEEE Internet of Things Journal}, vol.~10, no.~1, pp. 268--284, 2022.

\bibitem{lee2021creators}
L.-H. Lee, Z.~Lin, R.~Hu, Z.~Gong, A.~Kumar, T.~Li, S.~Li, and P.~Hui, ``When creators meet the metaverse: A survey on computational arts,'' \emph{arXiv preprint arXiv:2111.13486}, 2021.

\bibitem{huynh2023artificial}
T.~Huynh-The, Q.-V. Pham, X.-Q. Pham, T.~T. Nguyen, Z.~Han, and D.-S. Kim, ``Artificial intelligence for the metaverse: A survey,'' \emph{Engineering Applications of Artificial Intelligence}, vol. 117, p. 105581, 2023.

\bibitem{falchuk2018social}
B.~Falchuk, S.~Loeb, and R.~Neff, ``The social metaverse: Battle for privacy,'' \emph{IEEE Technology and Society Magazine}, vol.~37, no.~2, pp. 52--61, 2018.

\bibitem{ning2023survey}
H.~Ning, H.~Wang, Y.~Lin, W.~Wang, S.~Dhelim, F.~Farha, J.~Ding, and M.~Daneshmand, ``A survey on the metaverse: The state-of-the-art, technologies, applications, and challenges,'' \emph{IEEE Internet of Things Journal}, 2023.

\bibitem{heller2020reimagining}
B.~Heller, ``Reimagining reality: human rights and immersive technology,'' \emph{Harvard Carr Center Discussion Paper Series}, 2020.

\bibitem{online3}
\BIBentryALTinterwordspacing
Data availability and transparency bill 2022. [Online]. Available: \url{https://www.aph.gov.au/Parliamentary_Business/Bills_LEGislation/Bills_Search_Results/Result?bId=r6649}
\BIBentrySTDinterwordspacing

\bibitem{cheng2023towards}
R.~Cheng, S.~Chen, and B.~Han, ``Towards zero-trust security for the metaverse,'' \emph{IEEE Communications Magazine}, 2023.

\bibitem{10159439}
K.~Yang, Z.~Zhang, T.~Youliang, and J.~Ma, ``A secure authentication framework to guarantee the traceability of avatars in metaverse,'' \emph{IEEE Transactions on Information Forensics and Security}, vol.~18, pp. 3817--3832, 2023.

\bibitem{luo2021fa}
M.~Luo, J.~Cao, X.~Ma, X.~Zhang, and R.~He, ``Fa-gan: Face augmentation gan for deformation-invariant face recognition,'' \emph{IEEE Transactions on Information Forensics and Security}, vol.~16, pp. 2341--2355, 2021.

\bibitem{bernadelli2021dynamic}
C.~R. Bernadelli and P.~R. da~Silva, ``Dynamic time warping in iris biometric recognition process,'' \emph{IEEE Latin America Transactions}, vol.~19, no.~01, pp. 42--49, 2021.

\bibitem{chen2022query2set}
S.~Chen, Z.~Guo, X.~Li, and D.~Yang, ``Query2set: Single-to-multiple partial fingerprint recognition based on attention mechanism,'' \emph{IEEE Transactions on Information Forensics and Security}, vol.~17, pp. 1243--1253, 2022.

\bibitem{zhang2020learning}
Z.~Zhang, L.~Tran, F.~Liu, and X.~Liu, ``On learning disentangled representations for gait recognition,'' \emph{IEEE Transactions on Pattern Analysis and Machine Intelligence}, vol.~44, no.~1, pp. 345--360, 2020.

\bibitem{zhang2018continuous}
Y.~Zhang, W.~Hu, W.~Xu, C.~T. Chou, and J.~Hu, ``Continuous authentication using eye movement response of implicit visual stimuli,'' \emph{Proceedings of the ACM on Interactive, Mobile, Wearable and Ubiquitous Technologies}, vol.~1, no.~4, pp. 1--22, 2018.

\bibitem{ryu2022design}
J.~Ryu, S.~Son, J.~Lee, Y.~Park, and Y.~Park, ``Design of secure mutual authentication scheme for metaverse environments using blockchain,'' \emph{Ieee Access}, vol.~10, pp. 98\,944--98\,958, 2022.

\bibitem{shariff2023misogyny}
S.~Shariff, C.~Dietzel, K.~Macaulay, and S.~Sanabria, ``Misogyny in the metaverse: Leveraging policy and education to address technology-facilitated violence,'' in \emph{Cyberbullying and Online Harms}.\hskip 1em plus 0.5em minus 0.4em\relax Routledge, 2023, pp. 103--116.

\bibitem{allouzi2023adequate}
A.~AlLouzi and K.~Alomari, ``Adequate legal rules in settling metaverse disputes: Hybrid legal framework for metaverse dispute resolution (hlfmdr),'' \emph{International Journal of Data and Network Science}, vol.~7, no.~4, pp. 1627--1642, 2023.

\bibitem{wu2023financial}
J.~Wu, K.~Lin, D.~Lin, Z.~Zheng, H.~Huang, and Z.~Zheng, ``Financial crimes in web3-empowered metaverse: Taxonomy, countermeasures, and opportunities,'' \emph{IEEE Open Journal of the Computer Society}, vol.~4, pp. 37--49, 2023.

\bibitem{kostick2023ethical}
K.~Kostick-Quenet and V.~Rahimzadeh, ``Ethical hazards of health data governance in the metaverse,'' \emph{Nature machine intelligence}, pp. 1--3, 2023.

\bibitem{melzi2022overview}
P.~Melzi, C.~Rathgeb, R.~Tolosana, R.~Vera-Rodriguez, and C.~Busch, ``An overview of privacy-enhancing technologies in biometric recognition,'' \emph{arXiv preprint arXiv:2206.10465}, 2022.

\bibitem{bevsenic2022picking}
K.~Be{\v{s}}eni{\'c}, J.~Ahlberg, and I.~S. Pand{\v{z}}i{\'c}, ``Picking out the bad apples: unsupervised biometric data filtering for refined age estimation,'' \emph{The Visual Computer}, pp. 1--19, 2022.

\bibitem{Towards-privacy-EEG}
A.~Bidgoly, H.~Bidgoly, and Z.~Arezoumand, ``Towards a universal and privacy-preserving eeg-based authentication system,'' \emph{Sci Rep}, vol.~12, p. 2531, 2022.

\bibitem{Affective-EEG-Based}
T.~Wilaiprasitporn, A.~Ditthapron, K.~Matchaparn, T.~Tongbuasirilai, N.~Banluesombatkul, and E.~Chuangsuwanich, ``Affective eeg-based person identification using the deep learning approach,'' \emph{IEEE Transactions on Cognitive and Developmental Systems}, vol.~12, no.~3, pp. 486--496, 2020.

\bibitem{Fuzzy_Entropy_Metrics}
H.~Azami, P.~Li, S.~E. Arnold, J.~Escudero, and A.~Humeau-Heurtier, ``Fuzzy entropy metrics for the analysis of biomedical signals: Assessment and comparison,'' \emph{IEEE Access}, vol.~7, pp. 104\,833--104\,847, 2019.

\bibitem{head_authentication}
\BIBentryALTinterwordspacing
T.~Mustafa, R.~Matovu, A.~Serwadda, and N.~Muirhead, ``Unsure how to authenticate on your vr headset? come on, use your head!'' in \emph{Proceedings of the Fourth ACM International Workshop on Security and Privacy Analytics}, ser. IWSPA '18.\hskip 1em plus 0.5em minus 0.4em\relax New York, NY, USA: Association for Computing Machinery, 2018, p. 23–30. [Online]. Available: \url{https://doi.org/10.1145/3180445.3180450}
\BIBentrySTDinterwordspacing

\bibitem{random-forest}
T.~K. Ho, ``Random decision forests,'' in \emph{Proceedings of 3rd international conference on document analysis and recognition}, vol.~1.\hskip 1em plus 0.5em minus 0.4em\relax IEEE, 1995, pp. 278--282.

\bibitem{xg-boost}
\BIBentryALTinterwordspacing
T.~Chen and C.~Guestrin, ``{XGBoost}: A scalable tree boosting system,'' in \emph{Proceedings of the 22nd ACM SIGKDD International Conference on Knowledge Discovery and Data Mining}, ser. KDD '16.\hskip 1em plus 0.5em minus 0.4em\relax New York, NY, USA: ACM, 2016, pp. 785--794. [Online]. Available: \url{http://doi.acm.org/10.1145/2939672.2939785}
\BIBentrySTDinterwordspacing

\bibitem{lightgbm}
G.~Ke, Q.~Meng, T.~Finley, T.~Wang, W.~Chen, W.~Ma, Q.~Ye, and T.-Y. Liu, ``Lightgbm: A highly efficient gradient boosting decision tree,'' \emph{Advances in neural information processing systems}, vol.~30, pp. 3146--3154, 2017.

\bibitem{alain2023introduction}
M.~Alain, E.~Zerman, C.~Ozcinar, and G.~Valenzise, ``Introduction to immersive video technologies,'' in \emph{Immersive Video Technologies}.\hskip 1em plus 0.5em minus 0.4em\relax Elsevier, 2023, pp. 3--24.

\bibitem{ECG_12_electrode}
J.~Zheng, H.~Guo, and H.~Chu, ``A large scale 12-lead electrocardiogram database for arrhythmia study (version 1.0.0),'' \url{https://doi.org/10.13026/wgex-er52}, 2022, physioNet.

\bibitem{Auditory_Evoked-EEG}
N.~Abo~Alzahab, A.~Di~Iorio, L.~Apollonio, M.~Alshalak, A.~Gravina, L.~Antognoli, M.~Baldi, L.~Scalise, and B.~Alchalabi, ``Auditory evoked potential eeg-biometric dataset (version 1.0.0),'' \url{https://doi.org/10.13026/ps31-fc50}, 2021, physioNet.

\bibitem{apnea}
T.~Penzel, G.~Moody, R.~Mark, A.~Goldberger, and J.~Peter, ``The apnea-ecg database,'' \emph{Computers in Cardiology}, vol.~27, pp. 255--258, 2000.

\bibitem{body_sway}
J.~Streepey and S.~Dent, ``Body sway when standing and listening to music modified to reinforce virtual reality environment motion (version 1.0.0),'' \url{https://doi.org/10.13026/x32c-cz47}, 2021, physioNet.

\bibitem{gtzan}
G.~Tzanetakis and P.~Cook, ``Musical genre classification of audio signals,'' \emph{IEEE Transactions on Speech and Audio Processing}, vol.~10, no.~5, pp. 293--302, 2002.

\bibitem{norwegian_ecg}
B.~Singstad, ``Norwegian endurance athlete ecg database (version 1.0.0),'' \url{https://doi.org/10.13026/qpjf-gk87}, 2022, physioNet.

\bibitem{SHAP}
\BIBentryALTinterwordspacing
S.~M. Lundberg and S.-I. Lee, ``A unified approach to interpreting model predictions,'' in \emph{Advances in Neural Information Processing Systems 30}, I.~Guyon, U.~V. Luxburg, S.~Bengio, H.~Wallach, R.~Fergus, S.~Vishwanathan, and R.~Garnett, Eds.\hskip 1em plus 0.5em minus 0.4em\relax Curran Associates, Inc., 2017, pp. 4765--4774. [Online]. Available: \url{http://papers.nips.cc/paper/7062-a-unified-approach-to-interpreting-model-predictions.pdf}
\BIBentrySTDinterwordspacing

\end{thebibliography}

\end{document}